\documentclass[british]{article}
\usepackage[left=1in,top=1in,right=1in,bottom=1in,nohead,paperwidth=8.5in, paperheight=11in]{geometry} 
\usepackage{amsmath}
\usepackage{stmaryrd}
\usepackage{babel}
\usepackage{natbib} 
\usepackage{float}
\usepackage{rotating}
\usepackage{multirow}
\usepackage{algorithm}
\usepackage[titletoc, title]{appendix}
\usepackage{graphicx}
\usepackage{caption}
\usepackage{subcaption}

\title{\bf \large Multi-class Vector AutoRegressive Models for Multi-store Sales Data}
\author{Ines Wilms\thanks{Corresponding author ines.wilms@kuleuven.be}, Luca Barbaglia, and Christophe Croux \\ \textit{\small Faculty of Economics and Business, KU Leuven}}
\date{ }
\begin{document}
\maketitle

{\bf Abstract.}
Retailers use the Vector AutoRegressive (VAR) model as a standard tool to estimate  the effects of prices, promotions and sales in one product category on the sales of another product category.  Besides, these price, promotion and sales data are available for not just one store, but a whole chain of stores. 
We propose to study cross-category effects using a multi-class VAR model: we jointly estimate cross-category effects for several distinct but related VAR models, one for each store.
Our methodology encourages effects to be similar across stores, while still allowing for small differences between stores to account for store heterogeneity. Moreover, our estimator is sparse: unimportant effects are estimated as exactly zero, which facilitates the interpretation of the results.
A simulation study shows that the proposed multi-class estimator improves estimation accuracy by borrowing strength across classes.
Finally, we provide three visual tools showing
(i) the clustering of stores on identical cross-category effects, (ii) the networks of product categories and (iii) the similarity matrices of shared cross-category effects across stores.

\bigskip


\newpage

\section{Introduction \label{intro}}
Successful cross-category management requires retailers to understand ``cross-category demand effects", i.e. the effects of prices, promotions and sales of a certain product category on the sales (or demand) of another product category. 
The Vector AutoRegressive (VAR) model is ideal to measure such cross-category demand effects. In the $J$-dimensional VAR model of order $P$, the values of the $J$ price, promotion and sales time series are modeled as a function of their own past values, up to $P$ periods ago. 
As such, the VAR model accounts for time inertia in marketing spending and treats price and promotion variables as endogenous, thereby allowing feedback effects (e.g. \citealp{Dekimpe95}).
The relevance of cross-category analysis in the marketing literature is widely acknowledged (e.g. \citealp{Leeflang:12} and references therein).

To analyze cross-category demand effects, retailers typically prefer to work with store-level data (e.g. \citealp{Leeflang:12}). However, information on prices, promotions and sales is available not for only one store but typically for an entire chain of stores. 
A ``multi-class" VAR approach where we jointly estimate several distinct but related VAR models - one for each store  - is to be preferred to a standard VAR model.
Our multi-class approach  has several important advantages: 
(i) cross-category demand effects are expected to be \textit{similar} for the different stores since they belong to the same retail chain. We therefore encourage estimates to be similar among classes. As such, retailers can set a chain-wide marketing strategy for the shared dynamics across stores.
(ii) At the same time, we allow for differences between stores stemming from the \textit{heterogeneity} in shopping behavior at the different stores. As such, retailers can fine-tune their chain-wide strategy to accommodate store-specific effects.
(iii) By jointly estimating the multiple VAR models, we borrow strength across classes which results in \textit{improved estimation accuracy}, as will be illustrated by means of a simulation study. 
(iv) Our estimation method is ``\textit{sparse}" in the sense that many parameters are estimated as zero. Sparse estimation techniques have proven their worth in delivering highly interpretable VAR models in high dimensional settings, see amongst others \cite{Hsu08},  \cite{Abegaz13}, \cite{Basu15}, \cite{Davis15} and \cite{Gelper15}.


Sparse multi-class estimators have been recently introduced for graphical models \citep{Danaher14}, and regression models \citep{Kim09}. Our sparse multi-class estimator of the VAR model differs from the method of \cite{Kim09} in that 
(i) we consider a time series framework instead of a regression framework, 
(ii) we allow for a multivariate instead of a univariate response model for each class,  
(iii) we account for the correlation structure between the error terms of different equations of the VAR, and
(iv) we use the Smoothing Proximal Gradient  algorithm.

The remainder of this article is structured as follows.
Section \ref{method} introduces the multi-class VAR model, the corresponding estimator and algorithm. 
Simulation studies  in Section \ref{simulation} show the good performance of the proposed estimator in terms of estimation accuracy.
Section \ref{data}  presents the data and model for the multi-store sales application, Section \ref{case_results} discusses the  results.
Finally, Section \ref{conclusion} concludes.

\section{Multi-class VAR models \label{method}}

\subsection{Model and estimator}
Price, promotion and sales of several categories are available for each store (i.e. class) $1 \leq k \leq K$  over a certain period of time. Let $\textbf{y}_{t}^{(k)}=[y_{t,1}^{(k)},\ldots,y_{t,J}^{(k)}]'$ be a $J$-dimensional multivariate time series containing these price, promotion and sales data for store $k$ at a given point in time $1 \leq t \leq T$ where $T$ is the length of the time series.
The multi-class VAR model of order $P$ with $K$ classes and  $J$ time series is given by
\begin{equation}
\textbf{y}_{t}^{(k)}=\boldsymbol{B}_{1}^{(k)}\textbf{y}^{(k)}_{t-1}+\ldots+\boldsymbol{B}^{(k)}_{P}\textbf{y}^{(k)}_{t-P}+\textbf{e}^{(k)}_{t}\label{eq:VAR(P)}.
\end{equation}
The parameters $\boldsymbol{B}^{(k)}_{p}$, for $1 \leq p \leq P$ and $1 \leq k \leq K$,  are $J\times J$ matrices including all the autoregressive coefficients at lag $p$ for class $k$.
The $ij^{th}$ entry of $\boldsymbol{B}^{(k)}_{p}$ is denoted by $[\boldsymbol{B}^{(k)}_{p}]_{ij}:=\beta^{(k)}_{p,ij}$, for $1 \leq i,j\leq J$. 
This element measures the direct effect for class $k$ of time series $j$ on time series $i$ at lag $p$. As such,  we measure for each store $k$ the direct lagged effects of prices, promotions and sales in one category on the prices, promotions and sales of another category (including its own).
The error terms $\textbf{e}_{t}^{(k)}$ follow
a multivariate normal distribution $N_J({\bf 0}, \boldsymbol\Sigma^{(k)})$ and are independent over time.
We assume, without loss of generality, that all time series are mean centered such that no intercept is included.

We estimate the model parameters by penalized Generalized Least Squares. For ease of notation, 
rewrite model \eqref{eq:VAR(P)} in stacked form as 
\begin{equation}
\textbf{y}^{(k)}=\textbf{X}^{(k)}\boldsymbol\beta^{(k)}+\textbf{e}^{(k)}\label{eq:stacked_VAR(P)}, 
\end{equation}
where $\textbf{y}^{(k)}$ for  $1 \leq k \leq K$ is a $NJ$ vector stacking all $J$ time series, with $N=T-P$. For each class $k$, $\textbf{X}^{(k)}=I_{J}\varotimes \textbf{X}_{0}^{(k)}$, where the
$N\times JP$ matrix $\textbf{X}_{0}^{(k)}$ is defined as $\textbf{X}_{0}^{(k)}=[\underline{\bf y}_{1}^{(k)},\ldots,\underline{ \bf y}_{P}^{(k)}]$,
with $\underline{\bf y}_{p}^{(k)}$ being a $N\times J$ matrix collecting
the observations at lag $1 \leq p \leq P$ for the $J$ series
in the $k^{th}$ class. The symbol $\varotimes$ is the Kronecker product. Furthermore, $\boldsymbol\beta^{(k)}=[\beta_{1,11}^{(k)},\ldots,\beta_{P,JJ}^{(k)}]'$, and $\textbf{e}^{(k)}$ is the $NJ$ vector of stacked error components for each class $k$.

Given model \eqref{eq:stacked_VAR(P)}, we define the estimator $\widehat{\boldsymbol\beta}$ of the vector $\boldsymbol{\beta}=[\boldsymbol\beta^{(1)\prime},\ldots,\boldsymbol\beta^{(K)\prime}]'$ collecting the autoregressive parameters for all classes, as the minimizer of the following penalized Least Squares criterion 
\begin{equation}
\small
\widehat{\boldsymbol\beta}=\underset{\boldsymbol\beta}{\operatorname{argmin}}
\sum_{k=1}^{K}(\mathbf{y}^{(k)}-\mathbf{X}^{(k)}\boldsymbol{\beta}^{(k)})'(\mathbf{y}^{(k)}-\mathbf{X}^{(k)}\boldsymbol{\beta}^{(k)})+ \lambda_{1} P_1(\boldsymbol{\beta}) + \lambda_{2} P_2(\boldsymbol{\beta}) \label{FusedLasso},
\end{equation}
where $\lambda_{1}, \lambda_2>0$ are regularization parameters and $P_1(\boldsymbol{\beta})$, $P_2(\boldsymbol{\beta})$ are two penalty functions.

For the first penalty function, we take the $l_1$-penalty on the absolute value of the differences of corresponding autoregressive parameters across classes (e.g. \citealp{Tibshirani05}; \citealp{She10})
\begin{equation}
P_1(\boldsymbol{\beta}) = \sum_{k\neq k'}^{K}\sum_{i,j=1}^{J}\sum_{p=1}^{P}|\mathbf{\beta}_{p,ij}^{(k)}-\mathbf{\beta}_{p,ij}^{(k')}|. \label{P1beta}
\end{equation}
The aim of this penalty is to induce similarity across classes. 
The larger the value of $\lambda_1$, 
the more differences of corresponding autoregressive parameters will be set to zero. As a consequence, 
the more elements of $\widehat{{\bf B}}_p^{(1)},\ldots,\widehat{{\bf B}}_p^{(K)}$, for $1 \leq p \leq P$, will be identical across classes.
If $\lambda_1 \rightarrow \infty$, all corresponding autoregressive parameters across classes will be identical, hence the same VAR model is obtained for each class $k$. 
If $\lambda_1=0$, then each class $k$ has its own VAR model and there are no similarities across classes. 
Since some of the estimated autoregressive parameters will be identical for some classes, a ``clustering" of classes arises for each estimated autoregressive parameter, where all classes with the same estimated parameter value form a cluster.

For the second penalty function, we consider the $l_1$-penalty on the absolute value of the autoregressive parameters \citep{Tibshirani96}
\begin{equation}
P_2(\boldsymbol{\beta}) = \sum_{k=1}^{K}\sum_{i,j=1}^{J}\sum_{p=1}^{P}|\mathbf{\beta}_{p,ij}^{(k)}|. \label{P2beta}
\end{equation}
The aim of the second penalty is twofold. First, by adding this penalty to the objective function, estimation remains feasible if the number of parameters exceeds the time series length. Second, it induces sparsity in the estimated autoregressive parameters by setting some coefficients equal to zero. The larger the value of $\lambda_2$, the sparser the estimate of $\boldsymbol{\beta}$. The combination of the first and second penalty in the objective function in \eqref{FusedLasso} leads towards shared sparsity patterns across classes.

\medskip
We further improve the estimator \eqref{FusedLasso} by simultaneously estimating the correlation structure of the error terms. To this end, we include the inverse error covariance matrices  $\boldsymbol\Omega=[\boldsymbol\Omega^{(1)},\ldots,\boldsymbol\Omega^{(K)}]^{\prime}$ in the objective function, where $\boldsymbol{\Omega}^{(k)}=({\boldsymbol{\Sigma}^{(k)}})^{-1}$ for $1 \leq k \leq K$.
The $ij^{th}$ entry of $\boldsymbol{\Omega}^{(k)}$ is denoted by $[\boldsymbol{\Omega}^{(k)}]_{ij}:=\omega^{(k)}_{ij}$, for $1 \leq i,j\leq J$.
We define the \textit{multi-class} estimator as the minimizer of the following penalized Generalized Least Squares criterion 
\begin{multline}
\small
\mathrm{(\widehat{\boldsymbol\beta},\widehat{\boldsymbol\Omega})}=
\underset{\boldsymbol{\beta},\boldsymbol{\Omega}}{\operatorname{argmin}}\sum_{k=1}^{K}\left[(\mathbf{y}^{(k)}-\mathbf{X}^{(k)}\boldsymbol{\beta}^{(k)})'\mathbf{\widetilde{\Omega}}^{(k)}(\mathbf{y}^{(k)}-\mathbf{X}^{(k)}\boldsymbol{\beta}^{(k)})-NJ\log |\mathbf{\Omega}^{(k)}|\right]+\\
+ \lambda_{1} P_1(\boldsymbol{\beta}) + \lambda_{2} P_2(\boldsymbol{\beta})
+ \gamma_{1} P_1(\boldsymbol{\Omega}) + \gamma_{2} P_2(\boldsymbol{\Omega})\label{eq:GFLasso}, 
\end{multline}
where $\mathbf{\widetilde{\Omega}}^{(k)}=\mathbf{\Omega}^{(k)}\varotimes{\bf I}_{N}$, 
and $\gamma_{1}, \gamma_2>0$ are regularization parameters for the elements of the inverse error covariance matrices.
We use similar penalization on the elements of the inverse error covariance matrix as for the autoregressive parameters in equations \eqref{P1beta} and \eqref{P2beta}: 
\begin{equation}
P_1(\boldsymbol{\Omega}) = \sum_{k\neq k'}^{K}\sum_{i,j=1}^{J}|\omega_{ij}^{(k)}-\omega_{ij}^{(k')}| \ \ \ \ \text{and} \ \ \ \  
P_2(\boldsymbol{\Omega}) = \sum_{k=1}^{K}\sum_{i,j=1}^{J}|\mathbf{\omega}_{ij}^{(k)}|. \nonumber
\end{equation}
Hence, 
the larger the value of $\gamma_1$, the more elements of $\widehat{\boldsymbol \Omega}^{(1)},\ldots,\widehat{\boldsymbol \Omega}^{(K)}$ will be identical across classes.
The larger the value of 
$\gamma_2$, the sparser the estimate of $\boldsymbol{\Omega}$. Moreover, the penalty $P_2(\boldsymbol{\Omega})$ ensures that the estimate of the inverse error covariance matrix exists even when the number of parameters exceeds the time series length.
The elements of the inverse error covariance matrices $\boldsymbol\Omega^{(k)}$ have a natural interpretation as the partial correlations between the error terms of the $J$ equations for class $k$. If the $ij^{th}$ element of $\boldsymbol\Omega^{(k)}$ is equal to zero, this means that the error terms of equation $i$ and $j$ for class $k$ are independent given all the others.

\subsection{Algorithm}
This section provides technical details on the implementation of the algorithm. We iteratively solve the optimization problem in \eqref{eq:GFLasso} first considering  $\boldsymbol\beta$ conditional on $\boldsymbol\Omega$ and then  $\boldsymbol\Omega$ conditional on $\boldsymbol\beta$. The code of the algorithm is made available on http://feb.kuleuven.be/ines.wilms/software.
 
\medskip 
\noindent
{\bf Solving for $\boldsymbol\beta$ conditional on $\boldsymbol\Omega$.}
We build on \cite{Chen12} and extend 
their Smoothing Proximal Gradient (SPG) algorithm for sparse estimation of regression models.
The SPG algorithm optimizes a smooth approximation of the objective function (see also \citealp{Nesterov05}):
\begin{equation}
\widetilde{\boldsymbol{\beta}} = \underset{\boldsymbol{\beta}}{\operatorname{argmin}} \  g(\boldsymbol\beta)+h_{\mu}(\boldsymbol\beta)+\lambda_{2} P_2(\boldsymbol\beta), 
\label{smoothopt}
\end{equation}
where $g(\boldsymbol\beta)$ is the first term in the objective function in \eqref{eq:GFLasso} with $\boldsymbol{\Omega}$ kept constant, 
and we replace the term  $\lambda_1 P_1(\boldsymbol{\beta})$ with its smooth approximation
\[
h_{\mu}(\boldsymbol\beta)=\max_{||\boldsymbol\alpha||_{\infty}\leq1}\left(\boldsymbol\alpha'{\bf C}\boldsymbol\beta
- \dfrac{\mu}{2}||\boldsymbol\alpha||^{2}_{2} \right),
\]
with
$\mu>0$ a smoothing parameter, 
$\boldsymbol\alpha$ is a vector of auxiliary variables, and  
${\bf C}={\bf I}_P \varotimes {\bf \widetilde{C}}$ is the $(K-1)\frac{d}{2}\times d$ matrix,  with $d=\text{dim}(\boldsymbol{\beta})$, representing 
the pairs of coefficients that are coupled across classes.
One takes
$ {\bf \widetilde{C}} = [({\bf \widetilde{C}}_1 \varotimes {\bf I}_{J^2})^\prime , ({\bf \widetilde{C}}_2 \varotimes {\bf I}_{J^2})^\prime , \ldots , ({\bf \widetilde{C}}_{K-1} \varotimes {\bf I}_{J^2})^\prime]^\prime  $ with
\[
[{\bf \widetilde{C}}_k]_{ij}=\begin{cases}
\lambda_{1} & \text{if} \ j=i\\
-\lambda_{1} & \text{if} \ j=i+k\\
0 & \text{otherwise},
\end{cases}\ \ \ \ 
\]
for $ 1 \leq k \leq K-1, 1 \leq i  \leq K-k$ and  $1 \leq j \leq K$. The solution of the objective function in \eqref{smoothopt} is approximated using the FISTA algorithm \citep{Beck09}. 

Note that we choose the SPG algorithm  over other standard first-order methods since it has a theoretically faster convergence rate and it is more scalable to high-dimensional problems because of its lower per-iteration time complexity, see \cite{Chen12}.

\medskip

\noindent
{\it Selection of regularization parameters.} We use a two-dimensional grid of regularization parameters $\lambda_{1}, \lambda_{2}$ and search for the optimal ones minimizing the Bayesian Information Criterion
\begin{equation}
\text{BIC}_{\lambda_1, \lambda_2} = -2 g({\widetilde{\boldsymbol{\beta}}_{\lambda_1, \lambda_2}})+ \ df_{\lambda_1, \lambda_2} \operatorname{log}(N), \nonumber
\end{equation}
where ${\widetilde{\boldsymbol{\beta}}_{\lambda_1, \lambda_2}}$ is the estimator using the regularization parameters $\lambda_1,\lambda_2$ and  $df_{\lambda_1, \lambda_2}$ is the number of non-zero estimated components of ${\widetilde{\boldsymbol{\beta}}_{\lambda_1, \lambda_2}}$.
 
%

\bigskip

\noindent
{\bf Solving for $\boldsymbol\Omega$ conditional on $\boldsymbol\beta$.}
When $\boldsymbol\beta$ is fixed, the estimation in \eqref{eq:GFLasso} corresponds to the  Joint Graphical Lasso \citep{Danaher14} on the residuals ${\bf e}^{(k)}= {\bf y}^{(k)} - {\bf X}^{(k)}{\boldsymbol \beta}^{(k)}$, for $1 \leq k \leq K$. The Joint Graphical Lasso is computed using the fast Alternating Directions Method of Multipliers algorithm. The optimal values of the regularization parameters $\gamma_1$ and $\gamma_2$ are selected using the BIC (e.g. \citealp{Yuan07}).

\bigskip

\noindent
{\bf Starting value and convergence.} We start by taking $\boldsymbol{\Omega}^{(1)}=\ldots=\boldsymbol{\Omega}^{(K)}={\bf I}_{J}$, and then we solve for $\boldsymbol\beta$ conditional on $\boldsymbol\Omega$ and for $\boldsymbol\Omega$ conditional on $\boldsymbol\beta$. We iterate until the relative change in the value of the objective function in \eqref{eq:GFLasso} in two successive iterations is smaller than the tolerance value $\varepsilon=10^{-2}$. Convergence was reached in all simulation runs and  the real data example.

\section{Simulation Study \label{simulation}}
We compare the performance of the proposed ``multi-class" estimator, i.e. the solution of equation \eqref{eq:GFLasso},  with two alternative estimators: 
the ``Least Squares" (LS) estimator, where every VAR model is estimated separately for each class,
and the ``single-class" estimator, i.e. the solution of equation \eqref{eq:GFLasso} with $\lambda_{1} = \gamma_{1}= 0 $.
The Least Squares estimator is the standard unregularized estimator and 
the single-class estimator is the regularized estimator where the VAR is estimated sparsely but no similarities across classes are induced.

We simulate from a multi-class VAR of order $P=1$ with $K=15$ classes and $J=10$ time series. These dimensions are similar to the ones of our multi-store sales application. The data generating process for each class $k$ is:
\[
\textbf{y}_{t}^{(k)}=\textbf{B}^{(k)}_{1}\textbf{y}_{t-1}^{(k)}+\textbf{e}^{(k)}_{t},
\]
for $t=P+1,\ldots,T=100$ and $\textbf{e}^{(k)}_{t}$ follows a multivariate normal distribution with zero mean and covariance matrix $\boldsymbol\Sigma^{(k)}$.  
 
\paragraph{Simulation Designs.}
Table \ref{tab:DesignSimulations} in the Appendix reports the parameter values for the three simulation designs considered. 
In the ``Varying $\boldsymbol{\beta}$" Design, the autoregressive coefficients have the same sparsity structure, while the magnitude of the non-zero effects may vary across classes. 
We include dynamics among the different time series: time series two to ten lead time series one, while time series seven to ten lead time series six. 
Averaged across classes, the cross-effects are half the magnitude of the own lagged effects, and stationarity of the VAR is ensured.
The error covariance matrices are the same for all classes. 

In the ``Varying $\boldsymbol{\Sigma}$" Design, 
the inverse error covariance matrices are band matrices with the same sparsity pattern. 
The magnitude of the corresponding partial correlations varies across classes. 
In the ``Varying $\boldsymbol{\beta}\ \text{and} \ \boldsymbol{\Sigma}$" Design, the value of $\boldsymbol{\beta}$ is taken from the ``Varying $\boldsymbol{\beta}$" Design, the value of $\boldsymbol{\Sigma}$ is taken from the ``Varying $\boldsymbol{\Sigma}$" Design.

\paragraph{Performance measures.}
We compare the performance of the  estimators in terms of their estimation accuracy. Estimation accuracy is evaluated by the Mean Absolute Estimator Error, given by
\[
\mathrm{MAEE}=\dfrac{1}{R}\dfrac{1}{PKJ^{2}}\sum_{r=1}^{R}\sum_{k=1}^{K}\sum_{i,j=1}^{J}\sum_{p=1}^{P}|\widehat{\beta}_{p,ij,r}^{(k)}-\beta_{p,ij,r}^{(k)}|,
\]
where $\widehat\beta_{p,ij,r}^{(k)}$ is the estimate of $\beta_{p,ij}^{(k)}$ in simulation run $r$. We take $R=1000$ simulation runs.
%

\paragraph{Results.}
Table \ref{performance_simulations} reports the MAEE of the three estimators for the three simulation designs.  
The ``Varying $\boldsymbol{\beta}$" Design focuses on the performance in estimating the autoregressive coefficients when the errors of the different equations of each VAR are not correlated. 
The multi-class estimator attains a lower value of the MAEE than the single-class estimator: 0.083 versus 0.094 respectively. 
Accounting for the shared sparsity patterns across classes thus improves  estimation accuracy.
The difference in estimation accuracy is significant, as confirmed by a paired $t$-test with $p$-value$<0.01$. 
The two regularized estimators perform significantly better than the (unregularized) LS estimator. The multi-class estimator improves estimation accuracy by 33\% compared to the LS estimator. Since the number of parameters to be estimated is large compared to the time series length, the LS suffers from imprecise estimation accuracy.

\begin{table} 
	\caption{Simulated Mean Absolute Estimation Error of the three estimators for the three simulation designs.\label{performance_simulations}}
\centering
\begin{tabular}{lllccccccc} \hline
Design &&& LS   &&& Single-class   &&& Multi-class \\ \hline
Varying $\boldsymbol{\beta}$ Design 										&&&  0.124 &&& 0.094 &&& 0.083  \\
Varying $\boldsymbol{\Sigma}$ Design 										&&& 0.111 &&& 0.081   &&& 0.073  \\
Varying $\boldsymbol{\beta} \ \text{and} \  \boldsymbol{\Sigma}$ Design 	&&& 0.124 &&& 0.095 &&& 0.083 \\ \hline
\end{tabular}
\end{table}

The conclusions from the ``Varying $\boldsymbol{\Sigma}$" and ``Varying $\boldsymbol{\beta} \ \text{and} \ \boldsymbol{\Sigma}$" are similar: (i) the multi-class estimator attains the best estimation accuracy and significantly outperforms the other two estimators, (ii) 
the regularized estimators significantly outperform the  LS.
Since the \textit{multi-class} estimator attains the best overall estimation accuracy, we use this estimator to study the cross-category demand effects across multiple stores.

\section{Data and Model \label{data}}
We use data from Dominick's Finer Foods, a large Midwestern supermarket chain that operates in the Chicago metropolitan area. This database is well-established in the literature on cross-category analysis (e.g. \citealp{Wedel2004}; \citealp{Kamakura2007}; \citealp{Lang15}).
Weekly store-level scanner data are available on prices, promotions and sales.\footnote{For more information on the calculation of the prices, promotions and sales variables, see e.g. \cite{Srinivasan04}.}
We use this information to analyze cross-category demand effects between five categories involving drink items: Soft Drinks (SDR), Refrigerated Juices (RFJ), Beer (BER), Bottled Juices (BJC), and Frozen Juices (FRJ).   
These data are collected for $K=15$ stores over a period from January 1993 to July 1994, $T=77$ weeks in total.

Store-specific information is provided in Table \ref{storedemo} in the Appendix. Dominick adopts a price tier specific pricing strategy where each store belongs to one out of four price tier groups, i.e. Cub Fighter\footnote{Cub Fighters pursue a more aggressive pricing policy in comparison to the other price tiers.}, Low, Medium or High price tier. We consider 2 Cub Fighter, 2 Low price tier, 7 Medium price tier and 4 High price tier stores.
Table \ref{storedemo} also presents demographical characteristics of the consumers in each store's market area, namely
\verb|income|: logarithm of median income;
\verb|educ|: percentage of college graduates; 
\verb|ethnic|: percentage of blacks and hispanics;
\verb|hsizeavg|: average household size; and
\verb|hvalmean|: mean household value.

\medskip

We analyze cross-category demand effects in a multi-class VAR model consisting of $J=3\times5$ time series, for each of the $K=15$ classes and $T=77$ time points. The order of the VAR is selected using the BIC, and gives $P=1$. The estimated autoregressive parameters $\widehat{{\bf B}}_1^{(k)}$ from the multi-class VAR model in equation \eqref{eq:VAR(P)} capture the within- and cross-category effects for store $1 \leq k \leq K$. \textit{Within-category} effects are the effects of prices, promotion or sales on its own prices, promotions or sales. \textit{Cross-category} effects are the effects of prices, promotion or sales of a certain category on the prices, promotion or sales of another category.


\section{Multi-store Sales Application \label{case_results}}
%

As is common in the literature on cross-category analysis, we focus on the cross-category \textit{demand} effects, i.e. the effects of prices, promotion and sales of a certain category on the sales (or demand) of another category. 
A good understanding of these demand effects is valuable to retailers to better allocate their scarce marketing resources across categories.


Previous studies on cross-category demand effects either
(i) focus on a single store (e.g. \citealp{Leeflang:12}),
(ii) estimate separate models, one for each store (e.g. \citealp{Wedel2004}; \citealp{Gelper15})
, or
(iii) aggregate information from several stores (e.g. \citealp{Song:06}).
The first  two approaches do not exploit the similarity between stores belonging to the same retail chain. Moreover, separate store models are likely to produce more noisy, less stable estimates \citep{Lang15}. 
The third approach is likely to produce biased estimates since it ignores the fact that the data belong to different stores \citep{Kamakura2007}. Moreover, the differences between  stores are of interest to retailers wanting to set a store-specific strategy.
In contrast to these previous studies, we use the multi-class VAR approach from Section  \ref{method} and  discuss
(i) the clustering of stores on identical cross-category demand effects,
(ii) the product category networks, and
(iii) the similarity matrices of cross-category demand effects across stores.

\subsection{Store clustering} 
\begin{figure}
\centering
\begin{subfigure}{.5\textwidth}
  \centering
  \includegraphics[width=\linewidth]{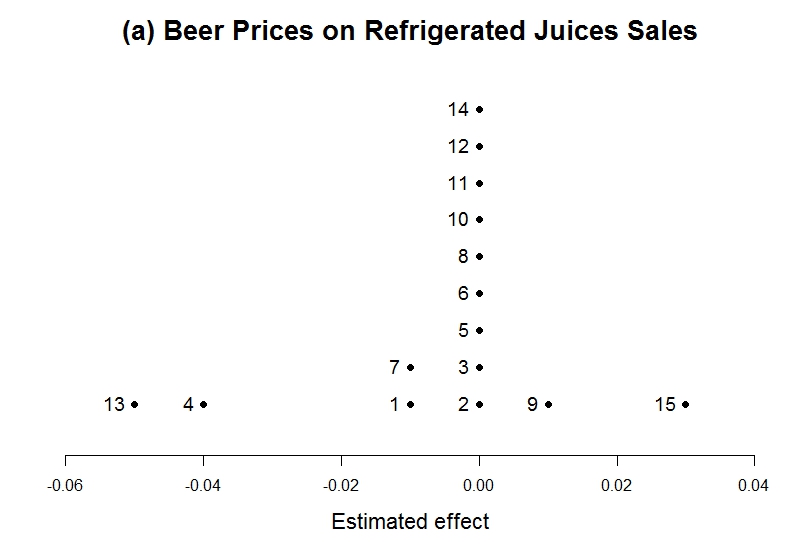}
  \label{Cluster:Price}
\end{subfigure}%
\begin{subfigure}{.5\textwidth}
  \centering
  \includegraphics[width=\linewidth]{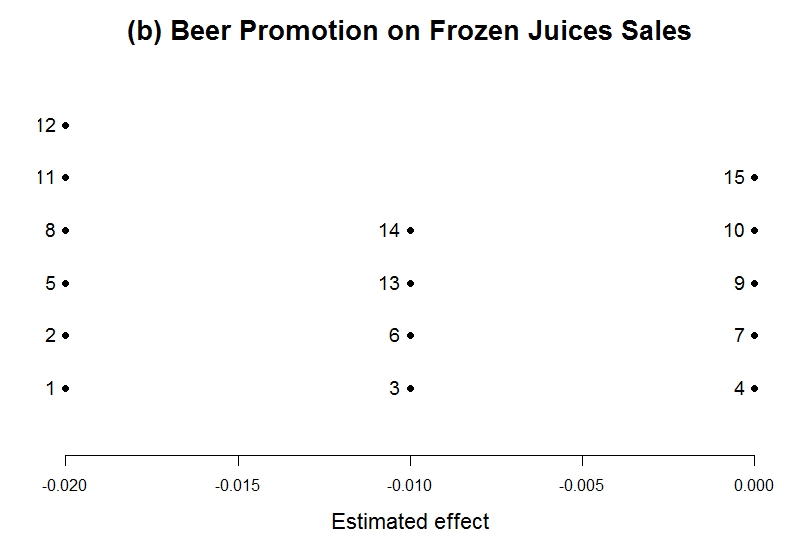}
  \label{Cluster:Promo}
\end{subfigure}
\begin{subfigure}{.5\textwidth}
  \centering
  \includegraphics[width=\linewidth]{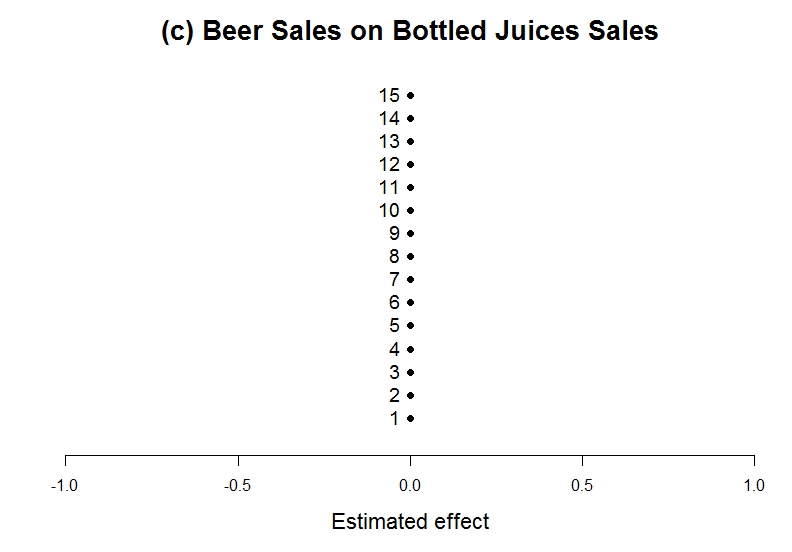}
  \label{Cluster:Sales}
\end{subfigure}%
  \caption{For each store (labeled from 1 to 15), we indicate the value of the estimated effect (horizontal axis) of (a) Beer prices on Refrigerated Juices sales, (b) Beer promotion on Frozen Juices sales, (c) Beer sales on Bottled Juices sales.} 
  \label{Cluster}
  \end{figure}

In Figure \ref{Cluster}, we consider three typical examples of estimated cross-category demand effects.
We indicate for each of the fifteen stores the value of the estimated cross-category demand effect (horizontal axis).
First, consider the estimated effects of Beer prices on Refrigerated Juices sales, see panel (a) of Figure \ref{Cluster}. 
For most stores, Refrigerated Juices sales are unresponsive (i.e. zero estimated effect) to a change in Dominick's Beer pricing.
The low-income, low-educated shoppers (cfr. low values of \verb|income| and \verb|educ| Table \ref{data}) at Store 9 and 15 are more subject to substitution effects: a price increase of Beer makes them substitute  Refrigerated Juices for Beer. 
In contrast, the small households with large homes (cfr. low value \verb|hsizeavg|, high value \verb|hvalmean|) at Store 13, or the high-income, high-educated shoppers at Store 1, 4 and 7 (cfr. high values of \verb|income| and \verb|educ|) are less vulnerable to substitution effects. 

Next, consider the estimated effects of Beer promotion on Frozen Juices sales, see panel (b) of Figure \ref{Cluster}. 
Frozen Juices sales are either unresponsive (i.e. zero estimated effect) or respond negatively to an increase in Dominick's Beer promotion intensity. This negative effect might be explained by substitutability: an increase in the promotion intensity of Beer, makes shoppers replace Frozen Juices by Beer. The low-income shoppers at Store 8, 12 and to a lesser extent Store 11 (cfr. low value \verb|income| Table \ref{data}) and the large households at Store 1, 2, and to lesser extend Store 5 (cfr. high value \verb|hsizeavg| Table \ref{data}) might be most vulnerable to this substitution effect.

Finally, consider the estimated effects of Beer sales on Bottled Juices sales, see panel (c) of Figure \ref{Cluster}. 
For all stores, Bottled Juices sales are unresponsive (i.e. zero estimated effect) to changes in Beer sales. 
Cross-category effects of sales on sales mainly occur due to the budget constraints: if consumers spend more on one category, they might, all else equal, spend less on another because they hit their budget constraint. 
Such effects are more likely to occur for categories where consumers spend a lot of their budget, and less likely to occur for categories where they spend less. Since consumers only spend, on average, 14\% and 10\% of their retail spending (in our data) on respectively Beer and Bottles Juices, this might explain why Bottled Juices sales are unresponsive to changes in Beer sales.

In sum, for each estimated cross-category demand effect, the multi-class estimator indicates how the different stores cluster together. 
Three possible scenarios can occur: 
(i) different clusters that vary in terms of \textit{sign and size} of the estimated effect (cfr. first example), 
(ii) different clusters that vary only in terms of \textit{size} of the estimated effect (cfr. second example), 
(iii) \textit{one cluster}: sign and size of the estimated effect is the same for all stores (cfr. third example).
In scenario (i), retailers should set out a store-specific strategy.
In scenario (ii), a store-specific strategy needs to be set only with respect to the expected degree of responsiveness of each store's market area.
Scenario (iii) allows retailers to set a chain-wide strategy.

\subsection{Product category networks \label{network}}
\begin{figure}
	\medskip
	\centering
	\includegraphics[scale=0.9]{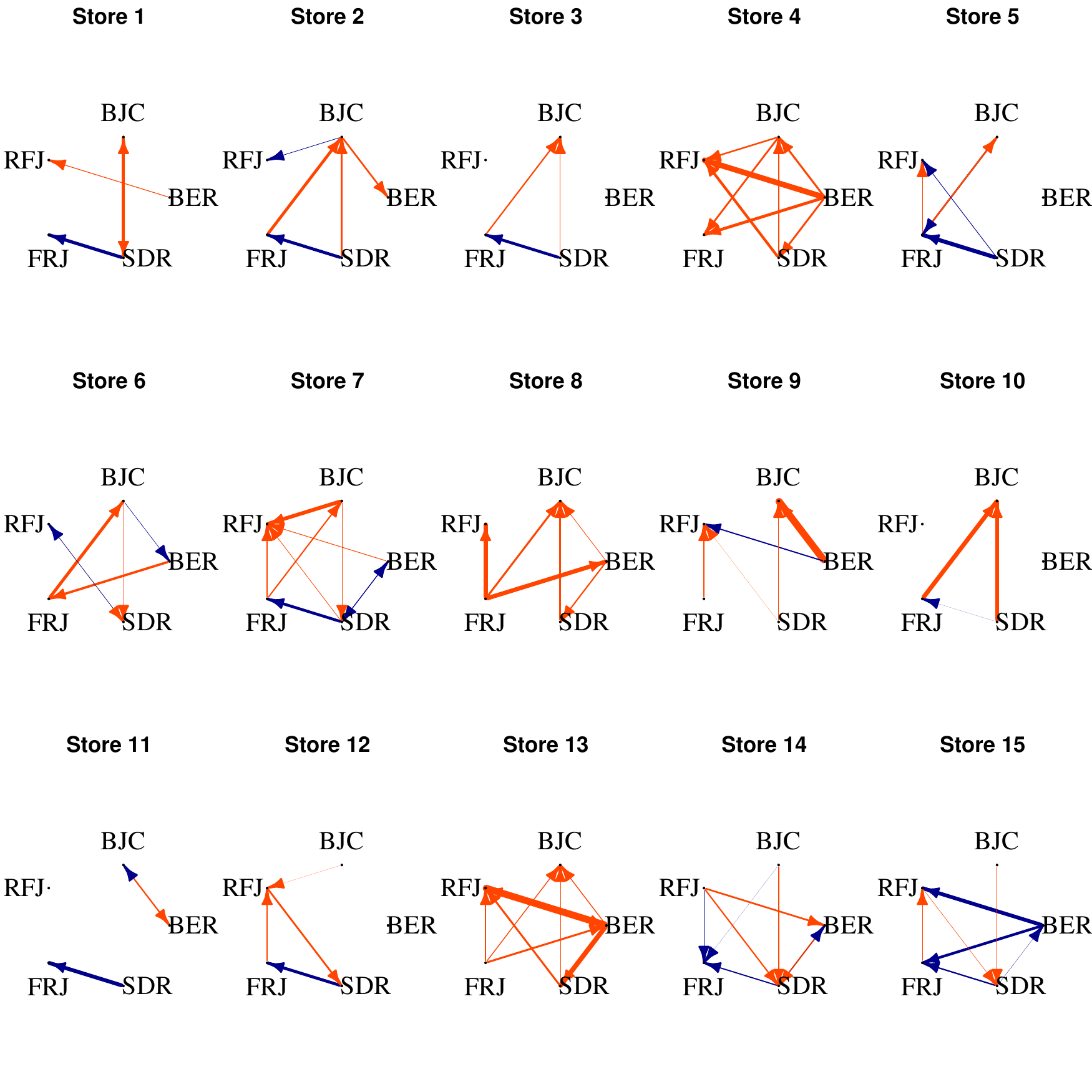}
		\caption{Product category network of \textit{prices on sales} for each of the 15 stores: a directed edge is drawn from one category to another if its prices influence sales in the other category. The edge width represents the
		magnitude of the effect. Positive effects are shown in blue,  negative effects in red (respectively dark and light gray on a gray scale). \label{PRICEonSLS}}
\end{figure}

We use a network analysis to get insights into the estimated cross-category demand effects.
The product category networks of \textit{prices on sales} are presented in Figure \ref{PRICEonSLS}. Fifteen networks are drawn, one for each store. The five product categories are the nodes of the networks. In each network, a directed edge is drawn from one category towards another if the multi-class estimator indicates, by giving a non-zero estimate, that prices in the former category have a direct influence on sales in the latter category. The edge width represents the effect size. Positive effects are shown in blue, negative effects in red.\footnote{On a gray scale: positive effects are shown in dark gray, negative effects in light gray.} Similar product category networks can be made for the effects of \textit{promotion on sales} and \textit{sales on sales}. For reasons of brevity, we only discuss the  network of \textit{prices on sales}. 


\paragraph{Asymmetry of cross-category demand effects.}
The cross-category effects of prices on sales are asymmetric. For example, a price increase in Soft Drinks makes consumers spend more on Frozen Juices as a compensation, (see the edge from SDR to FRJ for 9 stores in Figure \ref{PRICEonSLS}), yet a price increase on Frozen Juices does not affect the Soft Drinks sales. 
We typically find categories where consumers spend a lot of their budget, like Soft Drinks (i.e. 50\% of retail spending in our data), to be more influential than responsive: Soft Drinks has more outgoing than incoming edges in Figure \ref{PRICEonSLS} (i.e. 27  outgoing versus 14 incoming edges). Categories where consumers spend only a small fraction of their budget, like Bottled Juices (i.e. 10\% of retail spending in our data), are more responsive than influential: Bottled Juices has more incoming  than outgoing edges in Figure \ref{PRICEonSLS} (i.e. 21 incoming versus 15 outgoing edges). 
Similar conclusions regarding the asymmetry of cross-category effects of promotions on sales and sales on sales can be made.
This observed asymmetry is in line with previous research (e.g.  \citealp{Briesch2013}). 

An interesting finding concerns Soft Drinks at the High price tier Stores 12 to 15. For these stores, Soft Drinks is more responsive to price changes in other categories (1.75 incoming edges per store, on average) than for the other stores (0.64 incoming edges per store, on average). 
 Soft Drinks are less frequently consumed by High price tier shoppers \citep{Ogden11} and
less regularly purchased categories are typically expected to be more responsive to price changes in other categories, as is confirmed by our results.


\paragraph{Drivers of cross-category demand effects.}
We find considerably more negative cross-category effects  of prices on sales  than positive effects (71\% versus 21\%, on average, cfr. Figure \ref{PRICEonSLS}), but the positive effects are  about equally strong as the negative effects, on average.
Positive effects might be driven by the substitutability of the products belonging to different categories. 
Substitution effects occur between products that are perceived by consumers as substitute goods. For instance,  a price increase in Soft Drinks makes consumers purchase Frozen Juices  instead of Soft Drinks (for 9 stores in Figure \ref{PRICEonSLS}).
The somewhat more surprising negative effects might be explained by either reduced store traffic and/or budget constraints. 
Price increases might reduce store traffic and hence, lead towards lower overall sales (e.g.  \citealp{Wedel2004}). 
This especially holds for shoppers at the Cub Fighter and Low Price tier stores given their everyday-low-price positioning. At these stores, reduced store traffic is thus likely to be the main driver of the large number of negative effects of prices on sales.
Furthermore, price increases at one category might also constrain consumers' budget available for other categories, thereby leading towards lower sales of other categories and thus explaining the occurrence of negative cross-category effects of prices on sales. 

The results for Stores 1 to 3 require special attention.
Positive cross-category effects of prices on sales are much stronger than the negative effects. 
Substitutability is likely to be the main driver of the observed positive effects  at these Cub Fighter and Low price tier stores since their shoppers are typically more price-sensitive. Hence, the multi-class approach yields useful insights to retailers on how to accommodate their price tier specific retail strategy.

Finally, Store 4 and 13 show very specific cross-category effects of 
prices on sales: all observed effects are negative (cfr. Figure \ref{PRICEonSLS}). 
Their market areas are characterized by smaller household sizes, larger homes (low values of \verb|hsizeavg|, high values of \verb|hvalmean|, Table \ref{storedemo}), and either high-income, high-educated persons (high values \verb|income|, \verb|educ| for Store 4) or low percentage of blacks and hispanics (low value \verb|ethnic| for Store 13).  
These demographical variables are likely to reduce price sensitivity (\citealp{Mulhern98}),  making them less vulnerable to price substitution effects (i.e. positive price effects).  

\subsection{Similarity matrices}
\begin{figure}
\centering
\begin{subfigure}{.5\textwidth}
  \centering
  \includegraphics[width=\linewidth]{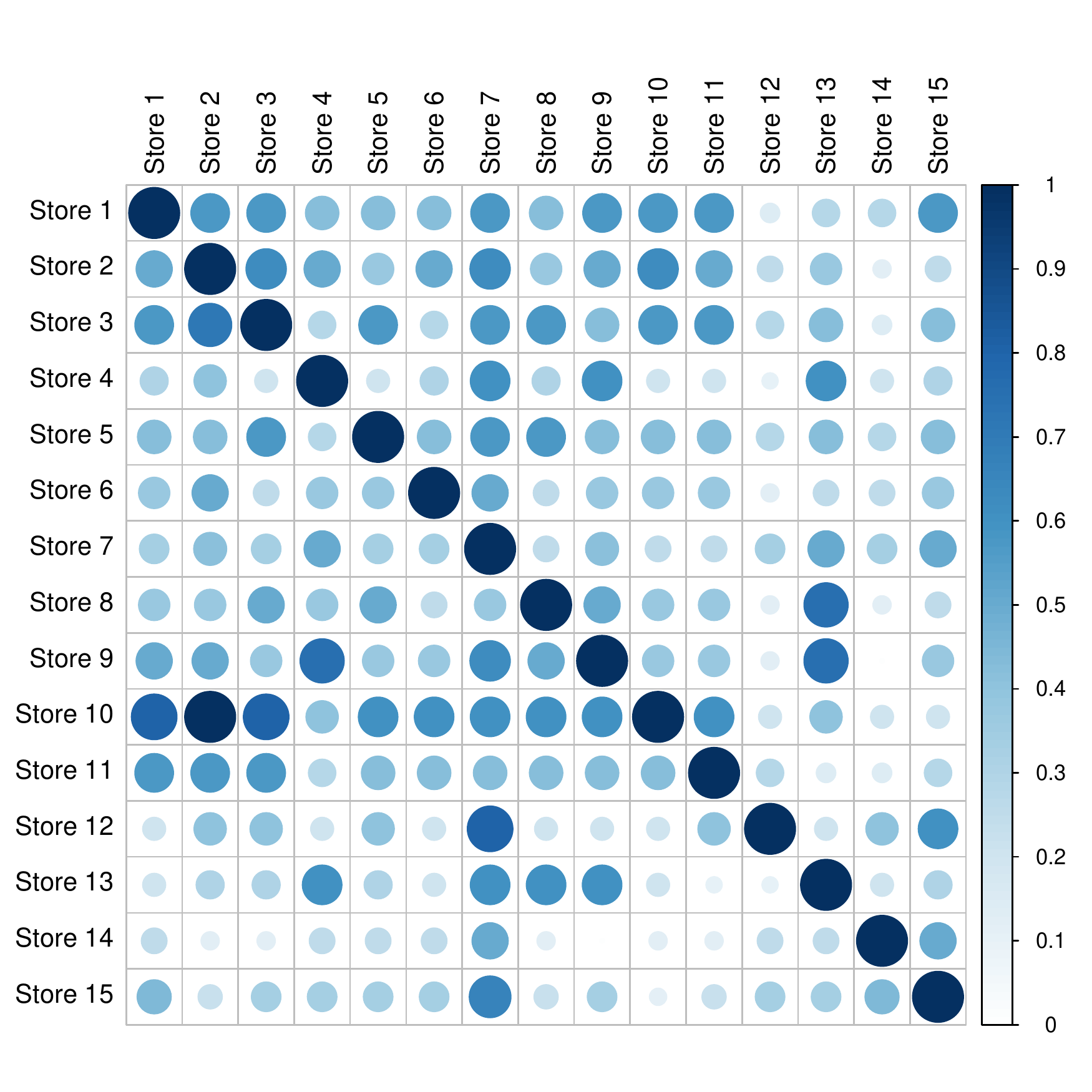}
  \caption{Prices on Sales}
  \label{NZ:PRICE}
\end{subfigure}%
\begin{subfigure}{.5\textwidth}
  \centering
  \includegraphics[width=\linewidth]{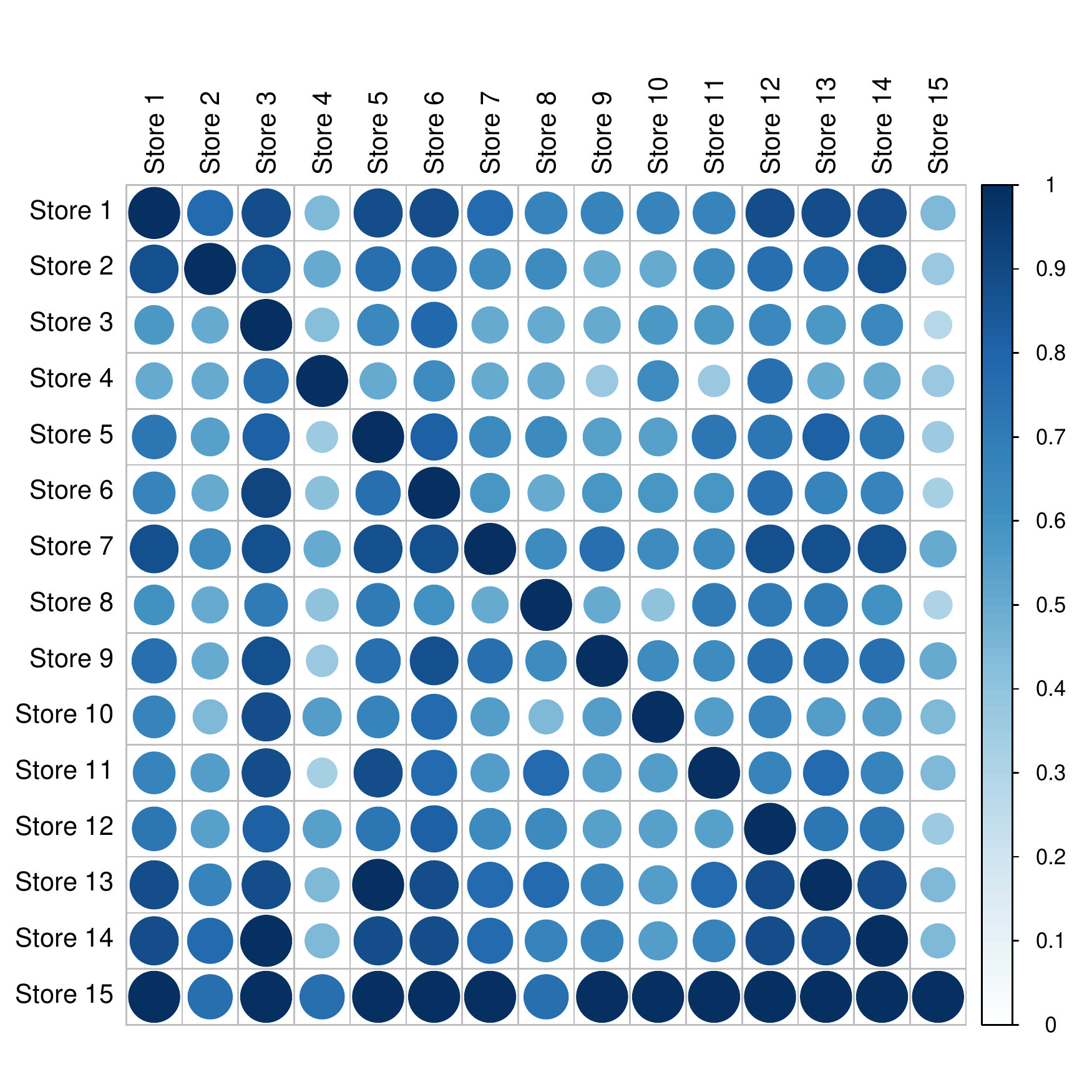}
  \caption{Promotions on Sales}
  \label{NZ:PROMO}
\end{subfigure}
\begin{subfigure}{.5\textwidth}
  \centering
  \includegraphics[width=\linewidth]{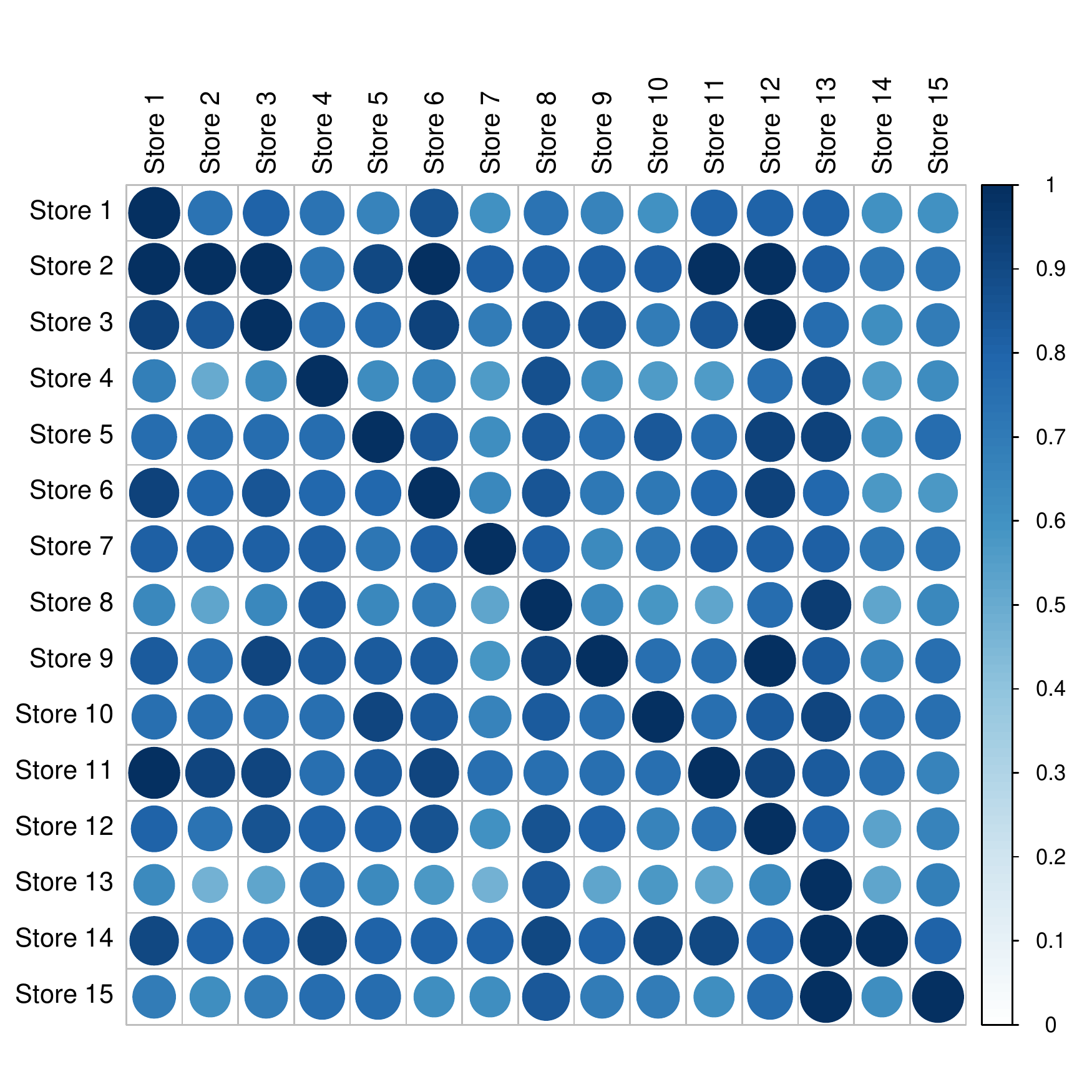}
  \caption{Sales on Sales}
  \label{NZ:SLS}
\end{subfigure}%
\caption{Similarity matrices. Each cell indicates the proportion of within- and cross-category effects of (a) prices on sales, (b) promotion on sales, and (c) sales on sales for store $i$ (row) that are also present for store $j$ (column). The darker and larger the circle, the higher the proportion.}
\label{NonZeroShared}
\end{figure}

We compare in Figure \ref{NonZeroShared}  the similarity matrices of shared  (within- and cross-category) demand effects across stores by computing for each pair of stores the proportion of shared non-zero effects   of \textit{prices on sales} (panel a), \textit{promotions on sales} (panel b), and \textit{sales on sales} (panel c). For instance, Store 10 and Store 1 share many \textit{prices on sales} effects, as indicated by the large size and dark color of the circle in the corresponding cell of panel (a): 80\% of the \textit{prices on sales} effects in Store 10 are also present for Store 1. 
In contrast, Store 10 and Store 15 share only a limited number of \textit{prices on sales} effects, as indicated by the small size and light color of the circle in the corresponding cell: only 20\% of \textit{prices on sales} effects in Store 10 are also present for Store 15.

The effects of \textit{sales on sales} and \textit{promotions on sales} show a considerably higher similarity across stores than the effects of \textit{prices on sales}. On average, stores share 76\% of \textit{sales on sales} , 67\% of \textit{promotions on sales}  and only 38\% of \textit{prices on sales} effects.
This low similarity of \textit{prices on sales} effects can be explained by Dominick's price tier specific pricing strategy (\citealp{Wedel2004}). 
Since prices at Dominick's stores are set differently according to the price tier type to which they belong (cfr. Table \ref{storedemo}),  \textit{prices on sales} effects are likely to vary considerable among stores. 
Dominick's promotional strategy, in contrast, is set more uniformly across stores, hence, explaining the higher similarity of \textit{promotions on sales} effects among stores.

Since Dominick adopts a price tier specific pricing strategy, we expect effects of \textit{prices on sales} to be more similar for stores belonging to the same price tier group than for stores of different price tier groups. This expectation is confirmed by our results: 
Cub Fighter stores share, on average, 76\% of \textit{prices on sales} effects,  whereas their shared effects with other price tier stores amounts to only 44\%, on average.
For Low price tier stores these percentages are 62\% versus 40\% respectively, for Medium price tier stores 51\% versus 39\%, for High price tier stores 49\% versus 30\%.

Looking at the shared \textit{prices on sales} effects of each pair of stores in Figure \ref{NonZeroShared} panel (a),  we find some results that can be explained by common market area demographics. 
For Store 10, for instance, 80\% of its \textit{prices on sales} effects are shared with Store 3. 
In terms of geographical proximity,  Store 10 is most closely located to Store 3.
Both stores operate in an area occupied by large households with small homes (cfr. high values of \verb|hsizeavg|, low values of \verb|hvalmean|, Table \ref{storedemo}).
Store heterogeneity stemming from store-level demographics is also found by, amongst others,  \cite{Chintagunta02},  and \cite{Sriram07}. 

\section{Conclusion \label{conclusion}}
This paper proposes a method for the joint estimation of multiple VAR models corresponding to distinct but related classes. By this joint estimation, we borrow strength across classes to estimate multiple VAR models that share certain characteristics. 
Our simulations show that this estimation approach results in a higher estimation accuracy. The proposed \textit{multi-class} estimator outperforms other estimators that do not encourage corresponding parameters across classes to be estimated identically.

We apply the multi-class VAR model to a multi-store sales application. 
The shared sales dynamics across stores allow retailers to design a chain-wide strategy that reflects the chain's image. 
The store-specific findings allow retailers to understand how each particular store responds to changes in its marketing mix. 
We provide visual tools helping to interpret the results of the multi-class VAR model.
They show (i) the store clustering, (ii) the product category networks and (iii) the similarity matrices of shared cross-category effects among stores.

The product category networks  visualize the estimated lagged effects captured in the autoregressive coefficient matrix. 
Alternatively, one could draw the product category networks based on the estimated impulse responses. The impulse response functions give the response of a certain time series to a unitary impulse in the error of another time series as a function of the  lag.  The network based analysis can then be extended by looking at, for instance, cumulative impulse responses.

Our multi-class VAR modeling approach is easily applicable to a variety of other settings. 
In biostatistics, the proposed methodology might be employed to analyze genetic data \citep{Abegaz13}. 
The time series contain gene expression measurements that are collected over time for a large number of genes. 
The classes are the treated patients and the controls.
The joint estimation could result in a more precise estimation of the gene regulatory networks.
In finance, one could study the differences and/or similarities in stock market dynamics among a set of connected financial institutions. The time series are stock market returns, the classes are the different financial institutions \citep{Diebold15}.
Another relevant application is the study of the dynamic relations among different pollutants across geographical areas \citep{Peng04}. Here the time series are the daily air pollutants levels, the classes are the difference stations for which the measurements are available.

\bigskip

\noindent
{\bf Acknowledgments.}
We gratefully acknowledge support from the FWO (Research Foundation Flanders, contract number 11N9913N) and from the GOA/12/014 project of the Research Fund KU Leuven.

\section*{Appendix: Additional Tables}
	\begin{table}[H]
		
		\caption {\small{Simulation designs of Multi-class VAR of order $P=1$ with $K=15$ classes and $J=10$ time series.} \label{tab:DesignSimulations}}
		\resizebox{0.75\textwidth}{!}{\begin{minipage}{\textwidth}
				\centering
				\begin{tabular}{l|ccccc} \hline
					Design & \multicolumn{2}{c}{$\boldsymbol\beta$}&&& $\boldsymbol\Sigma$ \\ \hline
					Varying $\boldsymbol{\beta}$ &  ${\bf B}_{1}^{(k)}=\begin{bmatrix}{\bf A}_{1}^{(k)} & {\bf A}_{2}^{(k)}\\
					0 & {\bf A}_{1}^{(k)}
					\end{bmatrix}$ &&&& $\boldsymbol\Sigma^{(1)}=\ldots=\boldsymbol\Sigma^{(K)}=\frac{1}{2}{\bf I}_J$\\
					&  & &&& \\
					& with \small{${\bf A}_{1}^{(k)}=\begin{bmatrix}
						0.5 & \eta^{(k)} & \eta^{(k)} & \eta^{(k)} & \eta^{(k)}\\
						0 & 0.5 & 0 & 0 & 0\\
						0 & 0 & 0.5 & 0 & 0\\
						0 & 0 & 0 & 0.5 & 0\\
						0 & 0 & 0 & 0 & 0.5
						\end{bmatrix}$} & \small{${\bf A}_{2}^{(k)}=\begin{bmatrix}
						\eta^{(k)} & \eta^{(k)} & \eta^{(k)} & \eta^{(k)} & \eta^{(k)}\\
						0 & 0 & 0 & 0 & 0\\
						0 & 0 & 0 & 0 & 0\\
						0 & 0 & 0 & 0 & 0\\
						0 & 0 & 0 & 0 & 0
						\end{bmatrix}$}  &&& \\ &  & &&& \\
					& where $\eta^{(k)}=\begin{cases}
					0.20 & if\ 1 \leq k \leq 5 \\
					0.25 & if\ 6 \leq k \leq 10 \\
					0.30 & if\ 11 \leq k \leq 15
					\end{cases}$  & &&& \\
					&&&&&\\
					&&&&&\\
					Varying $\boldsymbol{\Sigma}$ 	& ${\bf B}_{1}^{(1)}=\ldots={\bf B}_{1}^{(K)}=\begin{bmatrix}{\bf A}_{3} & {\bf A}_{4}\\
					0 & 0.5{\bf I}_{5}
					\end{bmatrix}$ &&&& $[\boldsymbol\Sigma^{(k)}]_{ij}=\frac{1}{2}{\rho_{(k)}}^{|i-j|}$\\
					&  & &&& \\		
					& with \small{${\bf A}_{3}=\begin{bmatrix}
						0.5 & 0.25 & 0.25 & 0.25 & 0.25\\
						0 & 0.5 & 0 & 0 & 0\\
						0 & 0 & 0.5 & 0 & 0\\
						0 & 0 & 0 & 0.5 & 0\\
						0 & 0 & 0 & 0 & 0.5
						\end{bmatrix}$} & \small{${\bf A}_{4}=\begin{bmatrix}
						0.25 & 0.25 & 0.25 & 0.25 & 0.25\\
						0 & 0 & 0 & 0 & 0\\
						0 & 0 & 0 & 0 & 0\\
						0 & 0 & 0 & 0 & 0\\
						0 & 0 & 0 & 0 & 0
						\end{bmatrix}$} &&& with $\rho_{(k)}=\begin{cases}
					0.05 & if\ 1 \leq k \leq 5 \\
					0.10 & if\ 6 \leq k \leq 10 \\
					0.15 & if\ 11 \leq k \leq 15
					\end{cases}$\\
					&&&&&\\
					&&&&&\\
					Varying $\boldsymbol{\beta} \  \text{and} \ \boldsymbol{\Sigma}$ 	& \multicolumn{2}{l}{$\boldsymbol\beta$ from ``Varying $\boldsymbol{\beta}$" Design} &&& $\boldsymbol\Sigma$ from ``Varying $\boldsymbol{\Sigma}$" Design\\	\hline					
				\end{tabular}
			\end{minipage} }
		\end{table}
		
\begin{table}[H]
	\caption{Store-specific price and demographical information. \label{storedemo}} 
	\centering
	\begin{tabular}{lllcccccccccc} \hline
		\verb|store| && \verb|price tier| && \verb|income| && \verb|educ| &&  \verb|ethnic| &&  \verb|hsizeavg|  &&  \verb|hvalmean|\\ \hline 
		1 && Cub Fighter	&& 10.716 && 0.178 && 0.105 && 3.110  && 120.134\\ 
		2 && Cub Fighter	&& 10.715 && 0.233 && 0.024 && 2.955  && 142.408\\ 
		3 && Low 			&& 10.597 && 0.095 && 0.035 && 2.770  && 97.501\\ 
		4 && Low 			&& 10.797 && 0.284 && 0.051 && 2.556  && 160.003\\ 
		5 && Medium 		&& 10.787 && 0.222 && 0.033 && 2.617  && 168.277\\ 
		6 && Medium 		&& 10.620 && 0.172 && 0.025 && 2.785  && 143.828\\ 
		7 && Medium 		&& 10.831 && 0.238 && 0.041 && 2.615  && 194.229\\ 
		8 && Medium 		&& 10.480 && 0.071 && 0.042 && 2.491  && 119.381\\ 
		9 && Medium 		&& 10.505 && 0.050 && 0.268 && 2.661  && 68.224\\ 
		10 && Medium 		&& 10.574 && 0.052 && 0.165 && 2.706  && 84.720\\ 
		11 && Medium 		&& 10.660 && 0.175 && 0.087 && 2.517  && 148.950\\ 
		12 && High 			&& 11.043 && 0.348 && 0.034 && 2.735  && 218.997\\ 
		13 && High 			&& 10.674 && 0.198 && 0.032 && 2.401  && 174.439\\ 
		14 && High 			&& 10.600 && 0.270 && 0.066 && 2.555  && 158.496\\ 
		15 && High 			&& 10.188 && 0.160 && 0.221 && 2.516  && 125.168\\ 
		\hline
	\end{tabular}
\end{table}


\bibliographystyle{asa}
\bibliography{GFLasso_ref}
	
\end{document}